\documentclass[aps,prc,twocolumn,showpacs,floatfix]{revtex4-1}

\usepackage{graphicx}% Include figure files
\usepackage{dcolumn}% Align table columns on decimal point
\usepackage{bm}% bold math
\usepackage{ulem}
\usepackage{color}

\newcommand{\al}{$\alpha$}

\newcommand{\raap}{($\alpha$,$\alpha'$)}
\newcommand{\rag}{($\alpha$,$\gamma$)}
\newcommand{\ran}{($\alpha$,n)}
\newcommand{\raneins}{($\alpha$,1n)}
\newcommand{\rann}{($\alpha$,2n)}
\newcommand{\rannn}{($\alpha$,3n)}
\newcommand{\rannnn}{($\alpha$,4n)}
\newcommand{\raxn}{($\alpha$,$x$n)}
\newcommand{\rap}{($\alpha$,$p$)}

\newcommand{\rgn}{($\gamma$,n)}
\newcommand{\rng}{(n,$\gamma$)}

\newcommand{\stot}{$\sigma_{\rm{reac}}$}
\newcommand{\sred}{$\sigma_{\rm{red}}$}

\newcommand{\seii}{$^{82}$Se}
\newcommand{\sevi}{$^{86}$Se}
\newcommand{\krix}{$^{89}$Kr}
\newcommand{\Nsv}{$N_A$$\left< \sigma v \right>$}

\newcommand{\rpro}{$r$-process}

\newcommand{\alpro}{$\alpha$-process}
\newcommand{\sfact}{S-factor}
\newcommand{\Ered}{$E_{\rm{red}}$}

\begin{document}

\title{
  Role of ($\alpha$,n) reactions under $r$-process conditions in
  neutrino-driven winds revisited
}

\author{Peter Mohr}
\email[Email: ]{WidmaierMohr@t-online.de}
\affiliation{
Diakonie-Klinikum, D-74523 Schw\"abisch Hall, Germany}
\affiliation{
Institute for Nuclear Research (ATOMKI), H-4001 Debrecen, Hungary}

\date{\today}

\begin{abstract}
\begin{description}
\item[Background] The astrophysical $r$-process occurs in an explosive
  astrophysical event under extremely neutron-rich conditions, leading to
  (n,$\gamma$)-($\gamma$,n) equilibrium along isotopic chains which peaks
  around neutron separation energies of a few MeV. Nuclei
  with larger $Z$ are usually produced by $\beta^-$-decay, but under certain
  conditions also $\alpha$-induced reactions may become relevant for the
  production of nuclei with $Z+2$.
\item[Purpose] The uncertainties of the reaction rates of these
  $\alpha$-induced reactions are discussed within the statistical model. As an
  example, $\alpha$-induced ($\alpha$,n) and $(\alpha$,$x$n) reaction cross
  sections for the neutron-rich $^{86}$Se nucleus are studied in detail.
\item[Method] In a first step, the relevance of ($\alpha$,n) and
  $(\alpha$,$x$n) reactions is analyzed. Next the uncertainties are determined
  from a variation of the $\alpha$-nucleus potential which is the all-dominant
  parameter for the astrophysical $Z \rightarrow Z+2$ reaction rate.
\item[Results] It is found that 
  the $r$-process flow towards nuclei with larger $Z$ is
  essentially influenced only by the $\alpha$-nucleus potential whereas the
  other ingredients of the statistical model play a very minor role. This
  finding is based on the fact that the flow towards larger $Z$ depends on the
  sum over all ($\alpha$,$x$n) cross sections which is practically identical
  to the total $\alpha$-induced reaction cross section. 
\item[Conclusions] $\alpha$-nucleus potentials play an important role under
  certain $r$-process conditions because the flow towards larger $Z$ depends
  sensitively on the total $\alpha$-induced reaction cross section. The
  uncertainty of the reaction rate is about a factor of two to three at higher
  temperatures and exceeds one order of magnitude at very low temperatures.
\end{description}
\end{abstract}

% 24.60.Dr 	Statistical compound-nucleus reactions
% 25.55.Ci 	Elastic and inelastic scattering
% 26.30.-k 	Nucleosynthesis in novae, supernovae and other explosive environments
% 26.30.Hj 	r-process
\maketitle

\section{Introduction}
\label{sec:intro}
The astrophysical \rpro\ is considered to be responsible for the
nucleosynthesis of about one half of the nuclei heavier than iron. In a
classical view, under the extremely neutron-rich \rpro\ conditions with
neutron densities above $10^{20}$/cm$^3$, matter is driven towards
neutron-rich nuclei by \rng\ reactions. An equilibrium is established between
\rng\ and \rgn\ reactions for nuclei with low neutron binding energies of the
order of a few MeV. Here the \rpro\ flow has to wait for the much slower
$\beta^-$-decay to proceed towards nuclei with larger Z (``waiting-point
approximation'') \cite{Cow91,Arn07,Thi11}.

It is obvious that the extreme conditions for the \rpro\ can only be achieved
in explosive scenarios. However, the astrophysical site(s) of the
\rpro\ are still under debate. The present study focuses onto the particular
conditions which are found in neutrino-driven winds above a nascent neutron
star or after the merging of two neutron stars. Here light \rpro\ elements may
be formed at high temperatures in a very short timescale of the order of
milliseconds. Under these conditions the $\beta^-$-decays may be too slow, and
thus nuclei with larger $Z$ can also be produced in a different way (e.g.,
\cite{Woo92,Ots00,Wan01,Tera01,Sumi01,Meyer02,Qian07,Qian08,Far10,Arc11,Arc13,Gor15,Mar15,Per16}).
The best candidate are \al -induced reactions in the so-called \alpro . These
reactions are often somewhat simplistic discussed as \ran\ reactions. However,
the following analysis will show that not only the \raneins , but also
\raxn\ reactions may contribute, and that the flow towards nuclei with larger
$Z$ is governed by the total \al -induced reaction cross section \stot .

Very recently, a sensitivity study on the theoretical uncertainties of
\ran\ reactions has been published by Pereira and Montes \cite{Per16}. For the
example of the \sevi \ran \krix\ reaction it was shown in \cite{Per16} that
the \raneins\ reaction rate is uncertain by at least an order of magnitude at
low temperatures below $T_9 \approx 3$ (with $T_9$ being the temperature in
$10^9$\,K) which is mainly based on the uncertainty which results from the
choice of the \al -nucleus potential. At higher temperatures above $T_9
\approx 5$ the uncertainty from the chosen \al -nucleus potential reduces to
about a factor of two to three, and the other ingredients of the statistical
model calculations lead to similar uncertainties (see Fig.~6 of
\cite{Per16}). Furthermore, it is noticed in \cite{Per16} that the widely used
code NON-SMOKER \cite{NONSMOKER} provides inclusive \ran\ cross sections and
rates, whereas the open-source code TALYS \cite{TALYS} calculates also
exclusive \raxn\ cross sections and rates. It is shown for the chosen example
of \sevi\ that the \raneins\ rate dominates at temperatures below $T_9 \approx
3$, whereas above $T_9 \approx 4$ the \rann\ rate exceeds the \raneins\ rate
significantly (Fig.~3 of \cite{Per16}). As \sevi\ is an unstable neutron-rich
nucleus four mass units ``east'' of the heaviest stable selenium isotope \seii
, the measurement of \al -induced cross sections for \sevi\ is extremely
difficult, and up to now experimental data are not available.

The present study fully agrees with the discussion of the astrophysical
scenario in \cite{Per16} and the conclusion on the importance of the \al
-nucleus potential. In addition to \cite{Per16}, this work
attempts to provide a better understanding of the
uncertainties of \al -induced reaction rates for the given astrophysical
\alpro\ scenario. For this purpose the following questions have to be
addressed. ($i$) What is the astrophysically relevant quantity? ($ii$) How
does this quantity depend on the underlying ingredients of the statistical
model? ($iii$) Is there a deeper understanding of the corresponding nuclear
physics? These questions will be answered in the following.

\section{The relevant reactions:\\ ($\alpha$,${\rm{n}}$) or
  ($\alpha$,$x{\rm{n}}$) ?} 
\label{sec:xn}
For simplicity and better readability, the following discussion uses the
example of \sevi\ which was also chosen in \cite{Per16}. It is pointed out in
\cite{Per16} that the \rng\ and \rgn\ reactions are faster than other
reactions by several orders of magnitude, thus leading to an equilibrium
isotopic distribution (e.g., within the selenium isotopic chain) which is a
function of temperature and neutron density. Let us assume that this isotopic
distribution peaks at \sevi . Following the approximation given in Eq.~(25) of
\cite{Arn07}, this corresponds e.g.\ to $T_9 = 3$ and $N_n \approx
10^{24}$/cm$^3$. Then the flow towards larger $Z$ may proceed via 
\sevi \raneins \krix , or in general via \sevi \raxn $^{90-x}$Kr (with $x =
1,\, 2,\, 3,\, {\rm{etc.}}$). Even the case $x = 0$, i.e.\ the \sevi \rag
$^{90}$Kr reaction, may contribute although the \rag\ cross section is
typically much smaller than the \raxn\ cross sections. 

As soon as any krypton isotope is made in this way, the fast \rng\ and
\rgn\ reactions drive krypton immediately towards $^{90}$Kr which has a
similar neutron separation energy as \sevi . This conclusion is completely
independent of the production by \raneins\ or \raxn\ reactions,
i.e.\ independent whether krypton is made as $^{86}$Kr or $^{90}$Kr. In
general, \rng\ rates increase with increasing positive \rng\ $Q$-value
towards less neutron-rich nuclei, i.e.\ towards stability. Thus,
e.g.\ $^{86}$Kr from the \sevi \rannnn $^{86}$Kr reaction is very efficiently
transmuted to $^{90}$Kr by a fast series of \rng\ reactions.

From the above arguments it becomes evident that the astrophysically relevant
quantity is the total production of Kr isotopes by \raxn\ reactions, i.e.\ the
sum over \raneins , \rann , \rannn , etc.\ including also the weak \rag\ cross
section. For the typical temperatures of the \alpro\ \cite{Per16}, the total
reaction cross section for the chosen example of \sevi\ is governed by the
\raneins\ channel. The reaction rate of the \rann\ channel contributes only
minor because of the negative $Q$-value of about $-4.5$\,MeV, and the reaction
rates of the \rannn\ and \rannnn\ channels are practically negligible.

\section{Total \al -induced reaction cross section \stot\ and
  the \al -nucleus potential} 
\label{sec:total}
The total reaction cross section \stot\ of \al -induced reactions is given by
the sum over all open reaction channels. In the case of neutron-rich
nuclei, any proton emission is highly suppressed because of the negative
$Q$-value. In the given example of \sevi\ the $Q$-value of the \rap\ reaction
is about $-7$\,MeV, and thus the astrophysical reaction rate \Nsv\ of the
\rap\ reaction remains negligibly small. As a consequence, the total reaction
cross section \stot\ is practically identical to the sum over the cross
sections of the neutron-emitting \raxn\ channels which has been identified as
the astrophysically relevant quantity in the previous
Sect.~\ref{sec:xn}. A minor contribution of inelastic \raap\ scattering to
the total reaction cross section \stot\ typically remains far below 
10\,\% at astrophysically relevant energies \cite{Orn16}.

\begin{figure}[htb]
\includegraphics[width=0.90\columnwidth,bbllx=40,bblly=25,bburx=340,bbury=555,clip=]{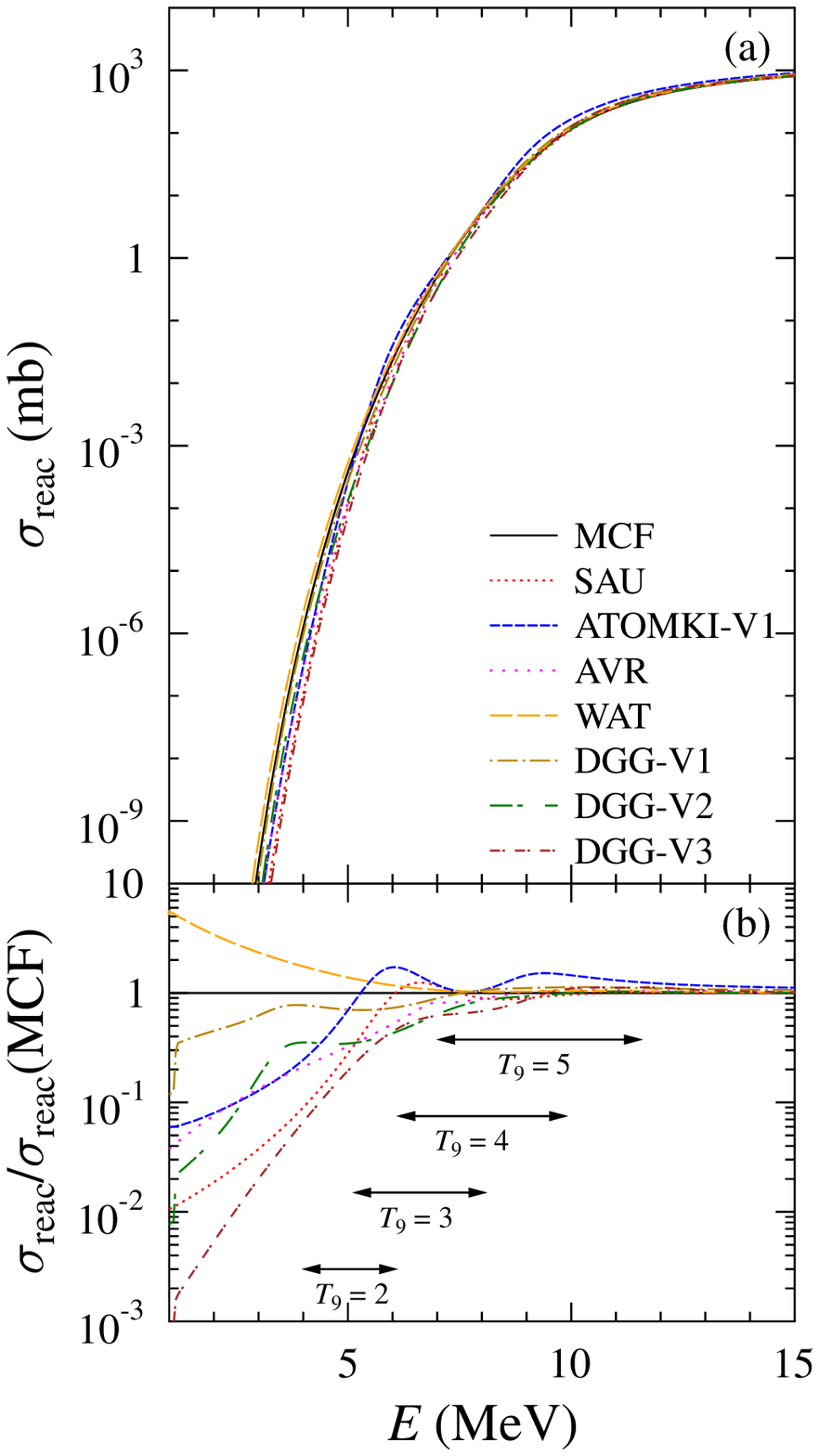}
\caption{
\label{fig:total}
(Color online)
Total reaction cross section \stot\ for \al -induced reactions on \sevi ,
calculated from different \al -nucleus potentials: MCF \cite{McF66}, SAU
\cite{Sau11}, ATOMKI-V1 \cite{Mohr13}, AVR \cite{Avr14}, WAT (TALYS default)
\cite{Wat58}, DGG: versions 1-3 from \cite{Dem02}. The upper part (a) shows
the cross sections which cover many orders of magnitude. The lower part (b)
shows the ratio normalized to the widely used McFadden/Satchler
potential. The Gamow window for temperatures $T_9 = 2 - 5$ is indicated by
horizontal arrows. Further discussion see text.
}
\end{figure}

The total reaction cross section \stot\ is related to the reflexion
coefficients $\eta_L$ by
\begin{equation}
\sigma_{\rm{reac}} = \frac{\pi}{k^2} \sum_L (2L+1) \, (1 - \eta_L^2)
\label{eq:stot}
\end{equation}
with the angular momentum $L$ and the wave number $k = \sqrt{2 \mu E}/\hbar$,
$E$ the energy in the center-of-mass system, and the reduced mass $\mu$ of
\al\ projectile and target. From a given \al -nucleus potential, 
the reflexion coefficients
$\eta_L$ and phase shifts $\delta_L$ can be calculated by solving the
Schr\"odinger equation. As the total reaction cross section \stot\ depends only
on the $\eta_L$ in Eq.~(\ref{eq:stot}), \stot\ depends only the the chosen \al
-nucleus potential, but not on the other ingredients of statistical model
calculations. A general behavior of the $\eta_L$ will be discussed in the
subsequent Sect.~\ref{sec:eta}. First I extend the sensitivity study of 
\cite{Per16} by including additional \al -nucleus potentials. In particular, I
include the recent many-parameter potential by Avrigeanu and coworkers (in the
version of \cite{Avr14}), the few-parameter ATOMKI-V1 potential
\cite{Mohr13}, and the modified McFadden/Satchler potential as suggested by
Sauerwein {\it et al.}\ \cite{Sau11}. I compare the results to the TALYS V1.6
default potential which is based on \cite{Wat58} and to the \al -nucleus
potentials of McFadden and Satchler \cite{McF66} and three different versions
suggested by Demetriou {\it et al.}\ \cite{Dem02}. The latest global \al
-nucleus potential by Su and Han \cite{Su15} has not been optimized for
energies below the Coulomb barrier, and it has been found in \cite{Orn16} that
it overestimates the experimental total reaction cross section \stot\ of
$^{64}$Zn at low energies significantly. In the present case of \sevi ,
\stot\ is also much higher by a about factor of 2 at 10\,MeV and more than one
order of magnitude at 5\,MeV. The results from the potential of \cite{Su15}
are thus omitted in Fig.~\ref{fig:total}.

Whereas the uncertainty study of \cite{Per16} discusses reaction rates, the
present work will compare the underlying cross sections. Here it will become
visible that at higher energies all predictions of \stot\ from the different
\al -nucleus potentials agree within about 10\,\% whereas dramatic
discrepancies are found at very low energies. For completeness it has to be
pointed out that the calculated cross sections in the present work are
calculated under laboratory conditions, i.e.\ without thermal excitations of
the \sevi\ target nucleus. However, for the case of \sevi , stellar
enhancement factors remain close to unity up to temperatures of about $T_9
\approx 5$ \cite{NONSMOKER}. 

The results for the different \al -nucleus potentials
\cite{Wat58,McF66,Sau11,Dem02,Avr14,Mohr13} are shown in
Fig.~\ref{fig:total}. Because the cross sections cover many orders of
magnitude, the lower part of Fig.~\ref{fig:total} shows in addition the ratio
normalized to the widely used McFadden/Satchler potential. It has been shown
recently that this potential provides an excellent description of \al -induced
reaction cross sections for relatively light nuclei in the $A \approx 20 - 50$
mass range \cite{Mohr15} whereas for heavier targets the McFadden/Satchler
potential tends to overestimate the cross sections in particular at low
energies below the Coulomb barrier. Note that the folding potential in the
three versions of the Demetriou {\it et.}\ \cite{Dem02} potential is
calculated within the TALYS code, whereas the folding potential of the
ATOMKI-V1 potential was derived from a 2-parameter Fermi distribution and
average parameters of neighboring nuclei given in \cite{Vri87}.

At higher energies above 15\,MeV the predictions from all \al -nucleus
potentials under study agree within about 10\,\%. This is an expected behavior
as will be shown in the next Sect.~\ref{sec:eta}. However, at lower energies
significant discrepancies can be found. Between 7 and 10\,MeV (corresponding
to the Gamow windows around $T_9 \approx 4 - 5$) the predictions show a
variation of about a factor of two to three. At even lower energies around
5\,MeV (corresponding to the Gamow window at $T_9 \approx 2$) the uncertainty
exceeds one order of magnitude. At very low energies, the range of predicted
\stot\ exceeds two orders of magnitude. It should be noted that \stot\ is very
small of the order of $10^{-20}$\,mb ($10^{-12}$\,mb) at $E = $ 2\,MeV
(3\,MeV). Fortunately, it is found that the numerical results from two
independent codes with slightly different default settings (TALYS which uses
ECIS \cite{ECIS} as subroutine for \stot , and A0 \cite{A0}) agree within a
few per cent down to such tiny cross sections; this discrepancy is further
reduced as soon as identical settings are chosen in both codes.

Summarizing Fig.~\ref{fig:total}, the upper part (a) shows the huge variation
of \stot\ with energy which results from the Coulomb barrier. The lower part
(b) visualizes that all \al -nucleus potentials under study agree very well at
energies above 15\,MeV whereas the range of predicted \stot\ increases
dramatically towards lower energies below the Coulomb barrier. The range of
predictions for the astrophysial reaction rate \Nsv\ can be estimated for
temperatures of $T_9 = 2 - 5$ from the marked Gamow windows. Compared to the
previous study \cite{Per16}, this range of predictions is somewhat increased
because three additional \al -nucleus potentials have been studied in this work.

The usual calculation of the Gamow window energies is based on the simplistic
assumption of a constant astrophysical \sfact\ which is not realistic for
heavy nuclei. Nevertheless, the Gamow window provides still a reasonable
estimate for the most relevant energy region for the astrophysical reaction
rate. Because the astrophysical \sfact\ typically decreases with increasing
energy for \al -induced reactions on heavy target nuclei, this leads to a
shift of the most effective energy towards lower energies by typically about
1\,MeV; for a detailed discussion of this shift, see \cite{Rau10}.

\section{General behavior of reflexion coefficients $\eta_L$}
\label{sec:eta}
The total reaction cross section \stot\ depends on the reflexion coefficients
$\eta_L$, see Eq.~(\ref{eq:stot}). It has already been discussed in detail
\cite{Mohr11,Mohr13b} that there is a general behavior of the $\eta_L$ at
energies above the Coulomb barrier. Partial waves with small angular momentum
$L$ (corresponding to small impact parameters or central collisions in a
semi-classical view) are practically fully absorbed ($\eta_L \approx 0$), and
partial waves with large $L$ (large impact parameters, peripheral
trajectories) are not absorbed ($\eta_L \approx 1$). 
\begin{figure}[t!]
\includegraphics[width=0.90\columnwidth,bbllx=25,bblly=25,bburx=360,bbury=790,clip=]{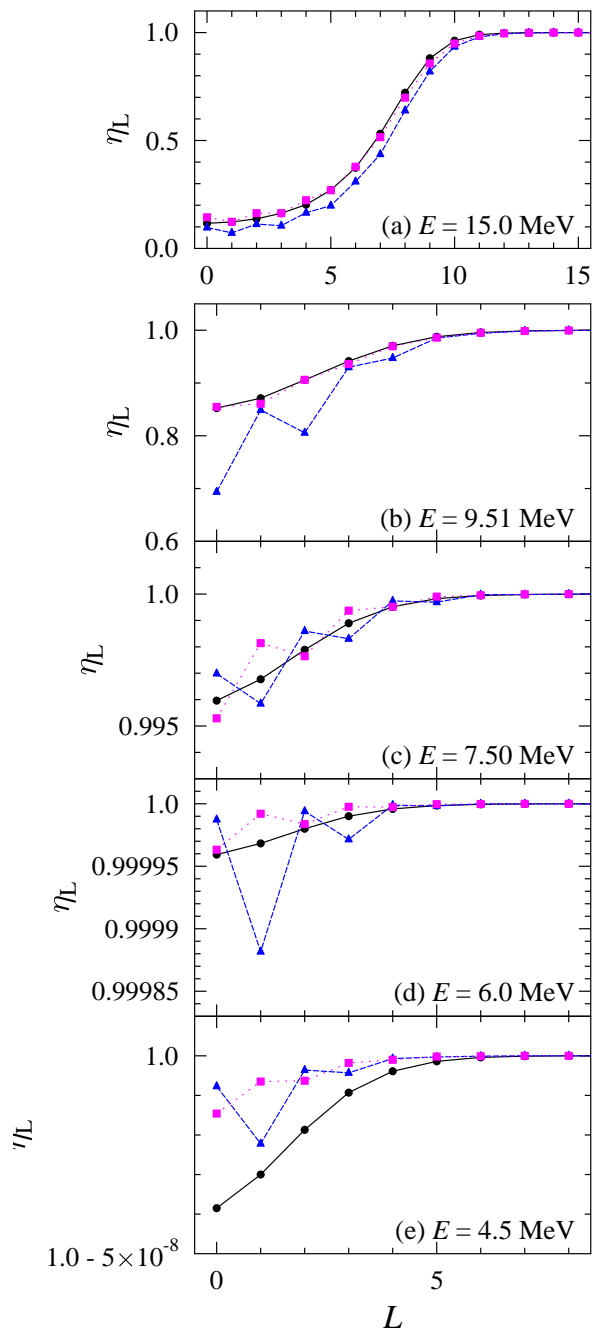}
\caption{
\label{fig:eta}
(Color online)
Reflexion coefficients $\eta_L$ at different energies above, around, and below
the Coulomb barrier, calculated from the potentials of McFadden/Satchler
\cite{McF66} (black circles), Avrigeanu {\it et al.}~\cite{Avr14} (magenta
squares), and ATOMKI-V1 \cite{Mohr13} (blue triangles). The corresponding
total reaction cross sections \stot\ from Eq.~(\ref{eq:stot}) are listed in
Table \ref{tab:stot}. At the highest energy of 15\,MeV (a), $\eta_L$ for
partial waves $L = 0 - 15$ are shown; here $\eta_L \approx 0$ for small $L$
and $\eta_L \approx 1$ for $L \gtrsim 10$. All potentials under study predict
this generic behavior above the Coulomb barrier. At lower energies (b-e) the
$\eta_L$ are shown for $L = 0 - 8$. Here only the $\eta_L$ for these few
partial waves deviate from unity and thus contribute to the sum for \stot\ in
Eq.~(\ref{eq:stot}). Now the calculated $\eta_L$ depend sensitively on the
properties of the \al -nucleus potentials. Note the extremely different scales
for $\eta_L$ in (a)-(e), reaching $1 - 5 \times 10^{-8}$ at the lowest energy
(e). The data points are connected by thin lines to guide 
the eye. Color codes and linestyles are identical to
Fig.~\ref{fig:total}. Further discussion see text.
}
\end{figure}

The transition from $\eta_L \approx 0$ to $\eta_L \approx 1$ happens within
few partial waves; consequently, the differences between any realistic
potentials are restricted to these few partial waves with $\eta_L \gg 0$ and
$\eta_L \ll 1$, and the resulting total reaction cross section \stot\ is
relatively well-defined as long as the chosen potential has a reasonable
radial range and a sufficient absorptive strength. In the chosen example of
\al +\sevi\ this behavior holds down to about 15\,MeV where the relevant
angular momentum number range is $5 \lesssim L \lesssim 10$ (see
Fig.~\ref{fig:eta}), leading to uncertainties for \stot\ of less than 10\,\%
above 15\,MeV.

The following discussion and presentation in Fig.~\ref{fig:eta} will focus on 
the widely used McFadden/Satchler potential (MCF) \cite{McF66}, 
the many-parameter potential by Avrigenau {\it et al.}\ (AVR) \cite{Avr14}, 
and the ATOMKI-V1 potential \cite{Mohr13}. The total reaction cross sections
\stot\ at the energies of Fig.~\ref{fig:eta} are listed in Table
\ref{tab:stot}.
\begin{table}[htb]
\caption{
\label{tab:stot}
Predictions of \stot\ from various global \al -nucleus potentials
\cite{Avr14,Mohr13,McF66}, corresponding to the $\eta_L$ shown in
Fig.~\ref{fig:eta}. 
}
\begin{tabular}{rcr@{$\times$}lr@{$\times$}lr@{$\times$}l}
\hline
\multicolumn{1}{c}{$E$ (MeV)} & &
\multicolumn{6}{c}{\stot\ (b)} \\
\multicolumn{2}{c}{} &
\multicolumn{2}{c}{AVR} & 
\multicolumn{2}{c}{ATOMKI-V1} & 
\multicolumn{2}{c}{MCF} \\
\multicolumn{2}{c}{} &
\multicolumn{2}{c}{Ref.~\cite{Avr14}} & 
\multicolumn{2}{c}{Ref.~\cite{Mohr13}} & 
\multicolumn{2}{c}{Ref.~\cite{McF66}} \\
\hline
15.0~ & ~ &
8.39 & $10^{-1}$    & 9.08 & $10^{-1}$   & 8.13 & $10^{-1}$ \\
9.51 & ~ &
6.95 & $10^{-2}$    & 9.98 & $10^{-2}$   & 6.59 & $10^{-2}$ \\
7.50 & ~ &
1.49 & $10^{-3}$    & 1.86 & $10^{-3}$   & 1.80 & $10^{-3}$ \\
6.0~ & ~ &
1.08 & $10^{-5}$    & 3.61 & $10^{-5}$   & 2.11 & $10^{-5}$ \\
4.5~ & ~ &
0.69 & $10^{-8}$    & 1.02 & $10^{-8}$   & 2.65 & $10^{-8}$ \\
\hline
\end{tabular}
\end{table}

At 15.0\,MeV total reaction cross sections \stot\ between 812 and 908\,mb are
found for the potentials
\cite{Wat58,McF66,Sau11,Dem02,Avr14,Mohr13}. Obviously, the largest deviations
of the $\eta_L$ from unity are found for the ATOMKI-V1 potential in
Fig.~\ref{fig:eta}(a), and thus, according to Eq.~(\ref{eq:stot}), the ATOMKI-V1
potential leads to a slightly larger \stot\ of 908\,mb whereas the MCF
(813\,mb) and the AVR (839\,mb) predictions are quite close to each
other. Overall, the deviations between the different potentials remain quite
limited with about 10\,\%.

The situation changes dramatically towards lower energies. Some further
energies were selected for illustration in Fig.~\ref{fig:eta} where
interesting properties can be seen for the cross section ratios in
Fig.~\ref{fig:total}. In general, at lower energies full absorption ($\eta_L
\approx 0$) is not reached for any partial wave. At energies far below the
Coulomb barrier all $\eta_L$ approach unity, and the total reaction cross
section is given by the tiny deviation of the $\eta_L$ from unity for very few
partial waves with small $L \lesssim 5$. These tiny
deviations depend sensitively on the chosen \al -nucleus potential.

\clearpage

At $E = 9.51$\,MeV the ATOMKI-V1 potential predicts a cross section which is a
factor of about 1.5 above the MCF and AVR predictions. Interestingly, this is
related to a relatively strong absorption of the even partial waves with $L =
0$ and $L = 2$ whereas the $\eta_L$ for the odd partial waves are almost
identical for all potentials, see Fig.~\ref{fig:eta}(b).

At $E = 7.50$\,MeV the three potentials under study provide almost identical
\stot . However, this agreement must be considered at random. The MCF
potential shows a smooth $L$ dependence, the AVR potential shows stronger
absorption for even $L$, and the ATOMKI-V1 potential favors absorption for odd
$L$, see Fig.~\ref{fig:eta}(c).

This odd-even staggering becomes more pronounced at $E = 6.0$\,MeV. Here the
strong absorption of the $L = 1$ partial wave leads to an ATOMKI-V1 cross
section which is about a factor of two above the MCF potential and a factor of
three above the AVR potential, see Fig.~\ref{fig:eta}(d).

At the lowest energy of $E = 4.5$\,MeV in Fig.~\ref{fig:eta}(e) \stot\ from
the MCF potential is about a factor of $2-3$ above the predictions from the
ATOMKI-V1 and AVR potentials. Towards even lower energies, the predictions of
ATOMKI-V1 and AVR agree surprisingly well whereas MCF predicts much larger
cross sections. Again, the relatively good agreement between ATOMKI-V1 and AVR
must be considered as accidential because of the discrepant underlying
$\eta_L$ from the ATOMKI-V1 and AVR potentials.

The odd-even staggering of the $\eta_L$ for the ATOMKI-V1 and AVR potentials
results directly from the numerical solution of the Schr\"odinger
equation. In both potentials the imaginary part is dominated by a surface
Woods-Saxon potential, i.e.\ absorption within a limited radial range. Thus,
the absorption becomes sensitive to the details of the wave function for each
partial wave. The different radius parameters ($R_S = 1.43$\,fm for ATOMKI-V1,
1.52\,fm for AVR; to be multiplied by $A_T^{1/3}$) lead to the different
behavior of the $\eta_L$ at the low energies in Fig.~\ref{fig:eta}. The
odd-even staggering is more pronounced for the ATOMKI-V1 potential with its
pure surface absorption, whereas the AVR potential includes also a small
volume Woods-Saxon imaginary potential at low energies. The odd-even
staggering does practically not appear for the MCF potential with its pure
volume Woods-Saxon imaginary part.

\section{Discussion}
\label{sec:disc}
\subsection{Cross sections}
\label{sec:cross}
The calculations of the total reaction cross section \stot\ in
Fig.~\ref{fig:total} and the underlying reflexion coefficients $\eta_L$ in
Fig.~\ref{fig:eta} show that \stot\ is relatively well defined within about a
factor of two to three down to about 8\,MeV. This covers the Gamow windows
above $T_9 \approx 4$. At lower energies down to about 5\,MeV, corresponding
to Gamow windows for the temperatures around $T_9 \approx 2-3$, the
uncertainty increases and reaches about one order of magnitude.

Recently, it has been found that so-called reduced cross sections and reduced
energies (as suggested in \cite{Gom05}) can be used to compare \al -induced
cross sections for many targets over a wide range of energies
\cite{Mohr15}. The reduced energy \Ered\ and the reduced cross section
\sred\ are defined by:
\begin{eqnarray}
E_{\rm{red}} & = & \frac{\bigl(A_P^{1/3}+A_T^{1/3}\bigr) E_{\rm{c.m.}}}{Z_P Z_T} \\
\sigma_{\rm{red}} & = & \frac{\sigma_{\rm{reac}}}{\bigl(A_P^{1/3}+A_T^{1/3}\bigr)^2}
\label{eq:red}
\end{eqnarray}

\begin{figure}[htb]
\includegraphics[width=0.90\columnwidth,clip=]{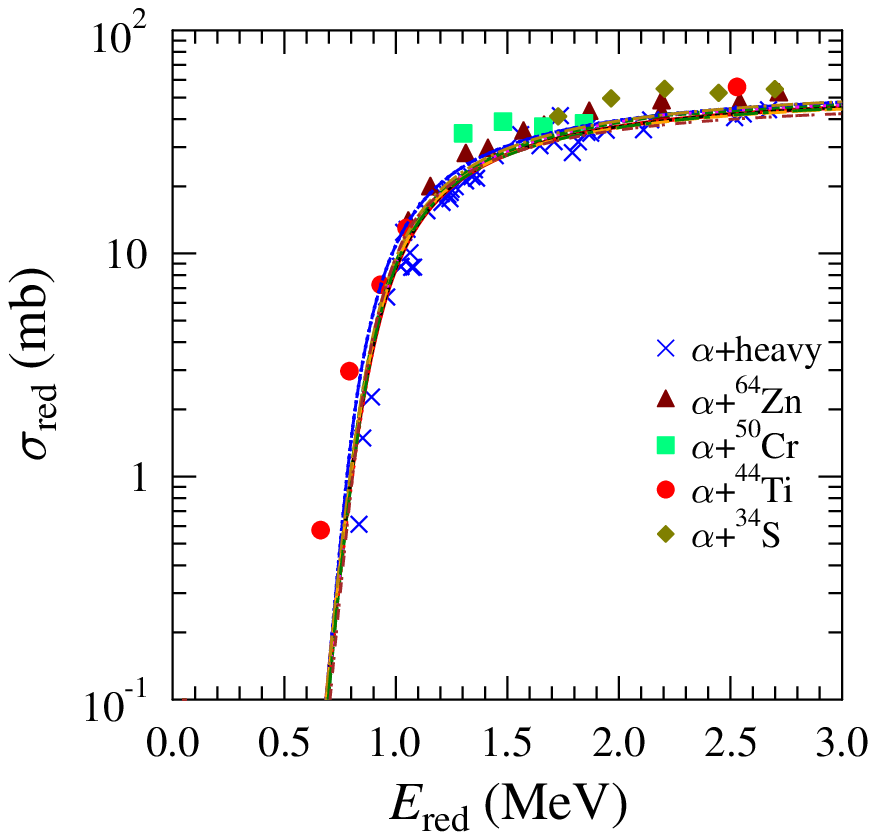}
\caption{
\label{fig:sred}
(Color online) 
Reduced cross sections \sred\ versus reduced energy \Ered\ for heavy ($A
\gtrsim 90$) and some lighter target nuclei (experimental data taken from
\cite{Mohr13,Mohr15}). The predictions for \sevi\ from the different potentials
\cite{Wat58,McF66,Sau11,Dem02,Avr14,Mohr13} are shown with lines; they are
relatively close (within a factor of two to three) for \Ered\ above $0.7$\,MeV
(corresponding to $E \approx 8$\,MeV), and thus the different lines appear
as almost identical in the logarithmic scale. Color codes and linestyles are
identical to Fig.~\ref{fig:total}.
}
\end{figure}

At reduced energies above $E_{\rm{red}} \approx 1.5$\,MeV all
nuclei show very similar \sred\ values of the order of $20 - 50$\,mb. Towards
lower energies, the Coulomb barrier leads to decreasing \sred . 
Fig.~\ref{fig:sred} shows experimental results for heavier ($A \gtrsim 90$)
targets with blue crosses; because the results remain very similar, the same
symbol has been chosen (for details see \cite{Mohr13,Mohr10}). The \sred\ for
lighter targets are somewhat larger than for heavier targets; experimental
results are shown for $^{64}$Zn, $^{50}$Cr, $^{44}$Ti, and $^{34}$S (taken
from \cite{Mohr15}). The predictions from eight different \al -nucleus
potentials \cite{Wat58,McF66,Sau11,Dem02,Avr14,Mohr13} for the neutron-rich
\sevi\ nucleus are shown as lines; these predictions fit nicely into the
general systematics in Fig.~\ref{fig:sred}.

The various potentials \cite{Wat58,McF66,Sau11,Dem02,Avr14,Mohr13} have been
determined from experimental data for stable nuclei. Obviously, an
extrapolation of the parameters is needed for the \al -nucleus potential of
the neutron-rich \sevi\ nucleus. The good agreement of the different
predictions at higher \Ered\ gives some confidence into this extrapolation but
unfortunately cannot further constrain the low-energy cross section and the
astrophysical reaction rate. Note that there is an approximate relation
between reduced energies and the Gamow window \cite{Mohr16}: $E_{{\rm{red}},0}
\approx 0.284\,{\rm{MeV}} \times T_9^{2/3}$. Consequently, the astrophysically
relevant range for the reduced energy \Ered\ is located below the shown range
of Fig.~\ref{fig:sred} which was chosen from the availability of experimental
data (taken from \cite{Mohr13,Mohr15}).

The above analysis of \al -induced reactions on \sevi\ and the role of
\ran\ and \raxn\ reactions can be extended to a broader range of target
nuclei. The general conclusions on the behavior of the $\eta_L$ will remain
valid, and the resulting uncertainties of \stot\ for a wider range of
targets will be quite similar to the chosen example of \sevi .

\subsection{Consequences for astrophysical reaction networks}
\label{sec:astro}
It has been shown above that the astrophysically relevant quantity for the
production of nuclei with $Z+2$ under \rpro\ conditions is the sum
over all \raxn\ cross sections which can be approximated by the total reaction
cross section \stot . As a consequence, the astrophysical reaction rate
depends only on the \al -nucleus potential, but is insensitive to the other
ingredients of the statistical model calculations. Although the other
ingredients do affect the branching ratios into the different \raneins , \rann ,
\rannn , etc.\ channels, they do not affect the total cross section \stot .

Compared to the recent study of Pereira and Montes \cite{Per16} where
uncertainties for the \raneins\ rate were estimated from all ingredients of the
statistical model, the present approach should in principle lead to smaller
uncertainties which are exclusively based on the uncertainty of the \al
-nucleus potential. However, such a reduction of uncertainties is only found
at very high temperatures where the \rann\ and \rannn\ channels contribute
significantly; such high temperatures exceed the typical range of the
\alpro\ as discussed in \cite{Per16}. At typical \alpro\ temperatures below
$T_9 \approx 3$, the \raneins\ channel is dominating the total reaction cross
section \stot . Consequently, also in \cite{Per16} the \al -nucleus
potential was identified as the dominating source of uncertainties. The
present study finds even a slightly increased uncertainty for the reaction
rate at low temperatures from the larger range of predictions from the three
additionally considered \al -nucleus potentials \cite{Sau11,Avr14,Mohr15}.

Extended astrophysical reaction networks should include all \raxn\ reactions
and their predicted rates e.g.\ from the TALYS code. As pointed out in
\cite{Per16}, the inclusive \ran\ rates from the NON-SMOKER code may induce
errors if they are considered as exclusive \raneins\ rates in such an extended
reaction network. However, this error remains small as long as the \rng
-\rgn\ equilibrium is established sufficiently fast and smears out the
produced isotopic distribution from the different \raxn\ reactions. On the
contrary, a significant error will occur as soon as a limited reaction network
which includes only the \raneins , but not the \raxn\ channels, is fed by the
exclusive \raneins\ rate e.g.\ from TALYS; here the flow towards nuclei with
larger $Z$ will be underestimated. Finally, it should be noted that the limited
reaction network will do a good job again using inclusive \ran\ rates of
NON-SMOKER or TALYS. 

Unfortunately, with the exception of \cite{Wan01}, none of
Refs.~\cite{Arn07,Thi11,Woo92,Ots00,Tera01,Sumi01,Meyer02,Qian07,Qian08,Far10,Arc11,Arc13,Gor15,Mar15}
states explicitly whether the chosen network considers the different
\raxn\ channels. The widely used REACLIB database \cite{REACLIB} contains for
the chosen example of \al +\sevi\ only the \rag , \ran , and \rap\ rates.
\raxn\ rates are not included in REACLIB. Thus, it seems very likely that most
of the \rpro\ network calculations use limited networks without explicit
consideration of the \raxn\ channels. As REACLIB recommends the
inclusive rates from NON-SMOKER, the final results should not be
affected dramatically by this limitation.

\section{Conclusions}
\label{sec:conc}
Very recently, Pereira and Montes \cite{Per16} have shown that
\raneins\ reaction rates depend sensitively on the chosen \al -nucleus
potential at low temperatures and show a weaker dependence on further
ingredients of the statistical model at higher temperatures. This finding is
correct for the exclusive \raneins\ rate. However, the present study shows that
the astrophysically relevant rate, i.e.\ the production of a nucleus with
$Z+2$ by \raxn\ reactions under \rpro\ conditions, is essentially defined by
the sum over all \raxn\ rates which is approximately given by the total \al
-induced reaction cross section \stot . This finding is based on the rapid
establishment of an equilibrium isotopic distribution by \rng\ and
\rgn\ reactions, which is independent of the particular \raxn\ production
reaction.  As the total reaction cross section \stot\ depends only on the
underlying \al -nucleus potential, but not on the other ingredients of the
statistical model, the uncertainty of the astrophysical rate can be well
estimated from the uncertainty of the \al -nucleus potential only.

It is found that the uncertainty of \stot\ at higher energies above 15\,MeV is
very small, whereas it increases dramatically towards lower energies.  This
leads to uncertainties of the reaction rate which are about a factor of two to
three for higher temperatures of $T_9 \approx 4-5$ and about one order of
magnitude for lower temperatures of $T_9 \approx 2-3$. At even lower
temperatures the uncertainty increases further. Compared to the study in
\cite{Per16}, the 
additional consideration of three recent \al -nucleus potentials of
\cite{Avr14,Sau11,Mohr15} slightly increases the range of predicted
\stot\ between 5 and 10\,MeV. The different predictions of \stot\ result from
different reflexion coefficients $\eta_L$ which depend sensitively on the
properties of the chosen \al -nucleus potential at low energies below the
Coulomb barrier. Interestingly, some cases have been identified where
discrepant predictions of $\eta_L$ lead to almost the same total reaction
cross section \stot\ which is given by sum over all contributing partial
waves. Any experimental test of the global \al -nucleus potentials
\cite{Wat58,McF66,Sau11,Dem02,Avr14,Mohr13} for nuclei with extreme
$N/Z$-ratio is very desirable. Such experiments may come in reach with the
upcoming radioactive ion beam facilities.

The astrophysical modeling of the \rpro\ in an extended network (including all
\raxn\ reaction channels) or in a limited network (with \raneins\ reactions
only) has to be consistent with the definition of exclusive \raxn\ (e.g.\ from
TALYS) or inclusive \ran\ cross sections and rates (as e.g.\ provided by
NON-SMOKER). The largest error occurs if a limited network is used in
combination with the exclusive \raneins\ rate (e.g.\ from TALYS); in this case
the \rpro\ flow towards larger $Z$ is underestimated because of the missing
contributions from the \raxn\ rates (with $x>1$). An extended network with the
inclusive \ran\ rate (e.g.\ from NON-SMOKER) for the exclusive
\raneins\ channel is not fully correct, but the \rng -\rgn\ equilibrium will
keep the resulting error relatively small.

\acknowledgments
I thank Zs.\ F\"ul\"op, Gy.\ Gy\"urky, and G.\ G.\ Kiss for many encouraging
discussions on \al -nucleus potentials, and J.\ Pereira and F.\ Montes for
their constructive criticisms. This work was supported by OTKA
(K108459 and K120666).

\end{document}